\documentclass[a4paper,11pt]{article}
\usepackage{jinstpub} 
\usepackage{lineno}

\title{\boldmath Neural Network Generalized Parton Distributions (NNGPD)}

\collaboration{\includegraphics[height=17mm]{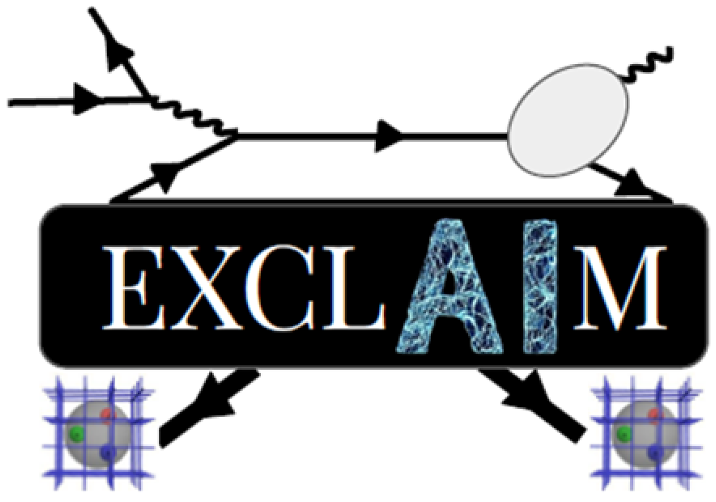}\\[6pt]
  EXCLusives via AI and ML \cite{Liuti:2024zkc}}

 \author[a, 1]{Zaki Panjsheeri,\note{Corresponding author.}}
\author[a]{Simonetta Liuti}
\affiliation[a]{University of Virginia Physics Department,\\
Charlottesville, VA 22904, USA}

\emailAdd{zap2nd@virginia.edu}

\abstract{Generalized parton distributions (GPDs) serve as indispensable tools for the exploration of proton structure. In this study, we offer a deep learning-assisted framework for the extraction of GPDs from experimental data and the results of ab-initio lattice quantum chromodynamics (LQCD).}


\begin{document}
\maketitle
\flushbottom

\section{Introduction}
\label{sec:intro}
GPDs parameterize off-forward matrix elements of QCD correlation functions; e.g., the leading-twist quark vector  GPDs $H$ and $E$ are defined as, 
\begin{eqnarray}
    W_{\Lambda'\Lambda}^{\gamma^{+}} &=& \frac{1}{2} \int \frac{dz^{-}}{2 \pi} e^{i x P^{+} z^{-}} \langle P', \Lambda' | \bar{\psi}(0, -\frac{1}{2} z^{-}, 0) \gamma^{+} \psi(0, \frac{1}{2} z^{-}, 0) | P, \Lambda \rangle_{z^{+} = 0, z_{T} = 0} \\
    &=& \frac{1}{2 P^{+}} \bar{u}(P', \Lambda') \Big(H^{q}(x, \xi, t; Q^{2}) \gamma^{+} + E^{q}(x, \xi, t; Q^{2}) \frac{i \sigma^{+ \nu} \Delta_{\nu}}{2M}\Big) , 
\end{eqnarray}
where a straight gauge link is chosen and appears trivially as unity in the correlation function. The GPDs are functions of $x$, the momentum fraction of the struck parton, $\xi$, the skewness which is related to $x_{Bj}$ (Bjorken x), $t$, the Mandelstam invariant defined as the square of the difference of the initial and final four-momentum of the proton, and $Q^{2}$, the hard scale. GPDs probe the partonic spatial structure of hadrons and parameterize the Ji and Jaffe-Manohar angular momentum sum rules \cite{Ji:1996ek, Radyushkin:1997ki, Jaffe:1989jz}. Extracting GPDs from experimental data would quantitatively answer the proton spin puzzle and offer a comprehensive spatial image of the proton (for the ExclAIm Collaboration's symbolic regression-based extraction of spatial structure from LQCD results, see Ref. \cite{Dotson:2025omi}). However, extracting GPDs from experimental data has proven to be a formidable challenge. Prior to engaging with actual experimental data, this is already evident from the formalism of deeply virtual exclusive scattering (DVES) \cite{Kriesten:2019jep, Kriesten:2020wcx}. The observables of DVES are not exactly GPDs, but rather the Compton form factors (CFFs), which are defined as the convolution of GPDs with Wilson coefficient functions, calculable in perturbative QCD in a factorized scenario \cite{Collins:1996fb, Ji:1998xh}. For example, the real and imaginary parts of the Compton form factor of the GPD $H$, corresponding to an unpolarized quark in an unpolarized proton, is defined at leading order perturbative QCD as, 
\begin{eqnarray}
\mathcal{H}_q(\xi,t)  = \Re e \mathcal{H}_q + i \, \Im m \mathcal{H}_q &=& P.V. \int_{-1}^{1} dx  \left( \frac{1}{x-\xi} + \frac{1}{x+\xi} \right)  H_q(x,\xi,t) \\
&& {- i \pi \, \big(H_q(\xi,\xi,t) - H_q(-\xi,\xi,t)\big)}.
\end{eqnarray}
The "first inverse problem" of GPDs, namely, the extraction of CFFs from DVES cross sections and asymmetries, is rife with difficulty due both to the formalism and to the quality of presently available data (see, e.g., the ExclAIm collaboration's statistically rigorous study of Jefferson Lab data in Ref. \cite{Adams:2024pxw}). Tabling for the moment this interesting first inverse problem, this study focuses on the "second inverse problem" of GPDs, which is the extraction of GPDs from CFFs. This second inverse problem is notably more prone to degenerate solutions than the extraction of parton distribution functions (PDFs) from deep inelastic scattering, given that the momentum fraction $x$ variable of GDPs is integrated over to give the CFFs, while the structure functions of PDFs maintain their functional dependence on $x$ as $x = x_{Bj}$ in inclusive processes. 

To address the challenges outlined above, we developed the NNGPD framework \cite{Xu:2026lko}, employing deep neural networks to construct high-dimensional, bias-controlled representations of GPDs constrained by QCD symmetries and experimental data.




\section{Machine learning framework}

\begin{figure}[htbp]
\centering
\includegraphics[width=.5\textwidth]{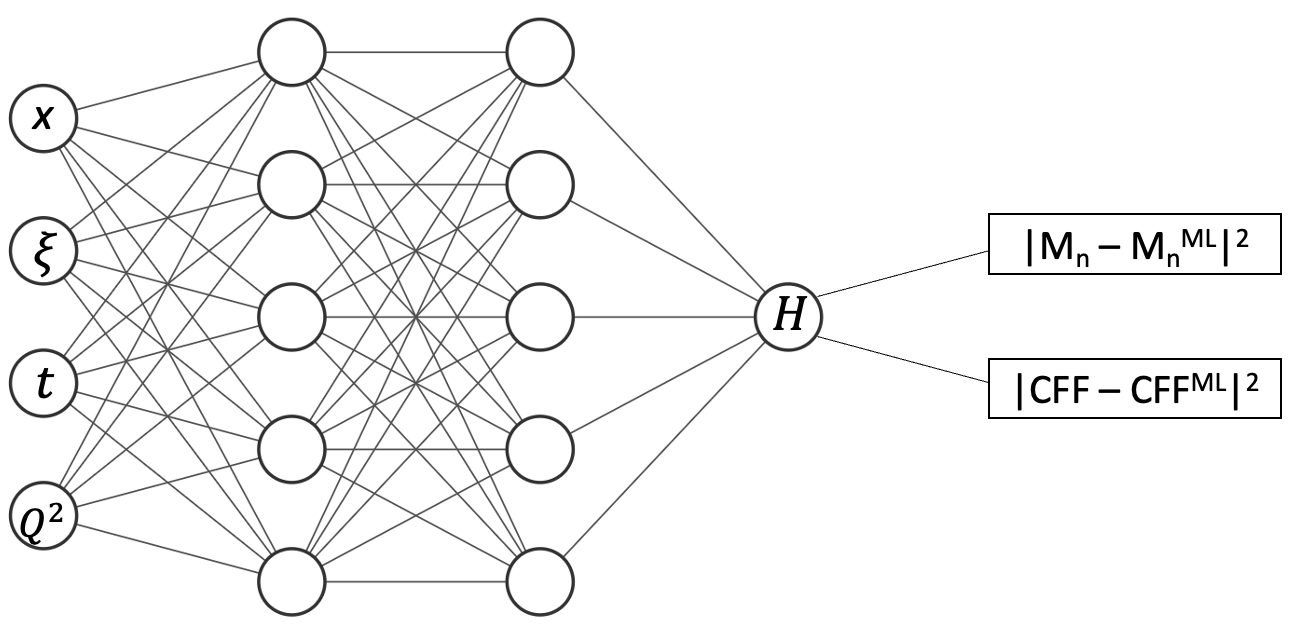}
\caption{Feed-forward neural network framework for NNGDP.\label{fig:ml_framework}}
\end{figure}

\noindent In addition to the CFFs, which are extracted from experiment, a robust further constraint on GPDs are the Mellin moments $M_n$, defined as $x$-weighted integrals of GPDs, 
\begin{eqnarray}
    M_{n}^{q}(\xi, t; Q^{2}) = \int_{-1}^{1}dx x^{n-1} H^{q}(x, \xi, t; Q^{2}). 
\end{eqnarray}
These Mellin moments can be shown to have a polynomial dependence on $\xi$ in terms of generalized form factors (GFFs), 
\begin{eqnarray}
    M_{n}^{q, H} = \int_{-1}^{1} dx x^{n-1} H_q(x, \xi, t) & = & \sum_{i=0,even}^{n-1} (2\xi)^i A^q_{n,i}(t) + \mod(n-1,2) (2\xi)^{n} C_{n}^q(t). 
\end{eqnarray}
Mellin moments and GFFs can be calculated through local insertions in the ab-initio framework of LQCD. We therefore propose a feedforward neural network framework that incorporates both the CFFs from experiment and the Mellin moments and generalized form factors calculated in LQCD, as shown in Figure \ref{fig:ml_framework}. Two different networks, one for the ``$+$" antisymmetric with respect to $x$ and another for the ``$-$" symmetric with respect to $x$ components of the GPDs, are defined through the losses as follows, taking into account that the ``$+$" distribution is constrained by the $n  \: \mathrm{even}$ moments, and the ``$-$" distribution is constrained by the $n \: \mathrm{odd}$ moments. 
The weights $w$ in the loss and the additional constraints under ``symmetry constraints" are explained in detail in Ref.  \cite{Xu:2026lko} within a Bayesian neural network (BNN).
The loss function $\mathcal{L}_{+}$ for $H^{+}_{q} \equiv H_{q} + H_{\bar{q}}$ is defined as,
\begin{eqnarray}
    \mathcal{L}_{+} = w_{\mathrm{CFF}} \mathcal{L}_{\mathrm{CFF}} + w_{\mathrm{MM}}^{+} \mathcal{L}_{\mathrm{MM}}^{+} + \: \mathrm{symmetry \: constraints}, 
\end{eqnarray}
where, 
\begin{eqnarray}
    \mathcal{L}_{\mathrm{CFF}} = | \mathrm{CFF} - \mathrm{CFF}^{\mathrm{ML}} |^{2}, 
\end{eqnarray}
meaning the difference between the ``ground truth" and ML-generated CFFs is taken during the loss minimization process, and a similar definition is given for the Mellin moments loss.  
The loss function $\mathcal{L}_{-}$ for the $H^{-}_{q} \equiv H_{q} - H_{\bar{q}}$ is defined similarly, using the odd/symmetric moments, but omitting the constraint from the CFFs.  
Considering both $\mathcal{L}_{+}$ and $\mathcal{L}_{-}$ then allows for the extraction of $H_{q}$ and $H_{\bar{q}}$. In a nutshell, the motivation for such a framework is that such an architecture is highly flexible and incorporates minimal physics bias. One can compare this framework to the scheme of the NNPDF collaboration where PDFs are modeled similarly as a neural network, multiplied by an ``endpoints regulating function," $x^{\alpha} (1 - x)^{\beta}$ (Ref. \cite{NNPDF:2014otw} and references therein). 

\subsection{Symmetry Constraints}
\begin{figure}[htbp]
\centering
\includegraphics[width=.7\textwidth]{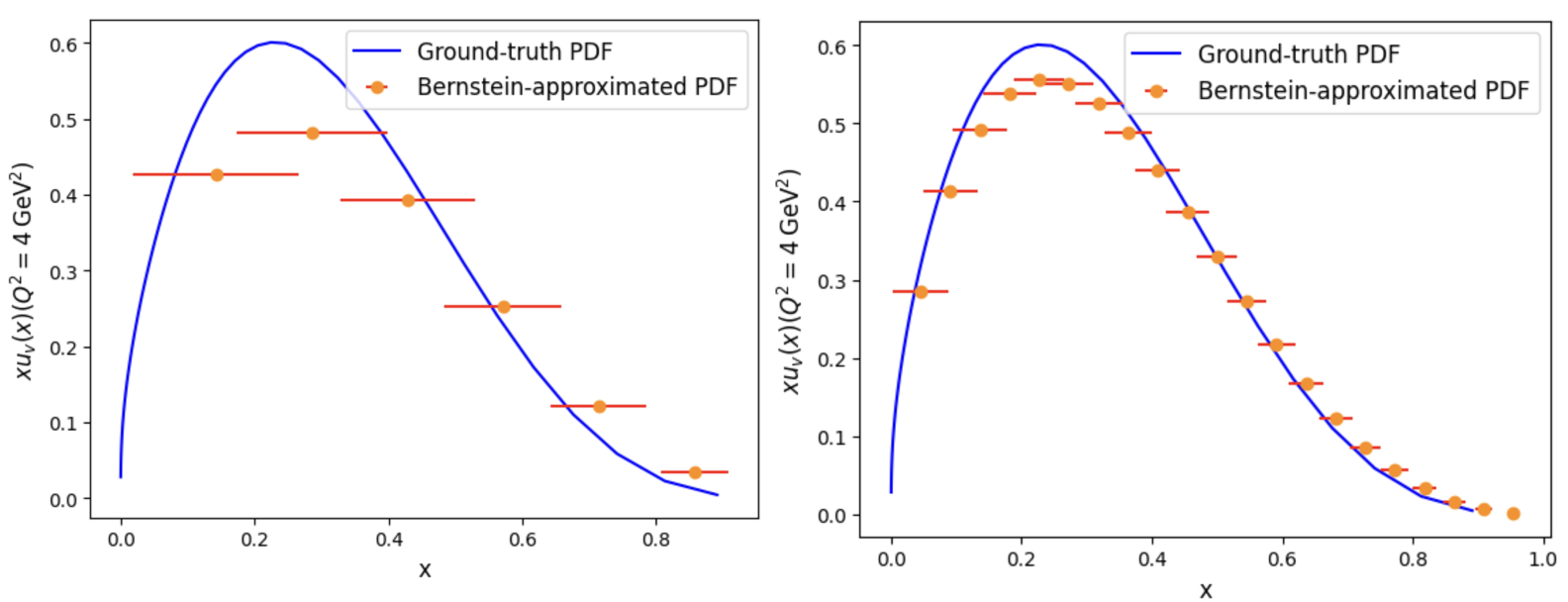}
\caption{Bernstein polynomial based extraction of a PDF using six moments (left) and twenty-one moments (right).\label{fig:Bernstein}}
\end{figure}

\noindent 
Traditional methods for extracting PDFs from moments employed Bernstein polynomials \cite{Yndurain:1977wz, Ahmad:2009fvg}. However, many moments are necessary to extract a large range in $x$, and it is especially difficult to reach low $x$, as demonstrated in an example with a typical valence PDF form in Figure \ref{fig:Bernstein}: notice that even with more than twenty moments, which is at present nonexistent in LQCD calculations, not much in the way of low $x$ can be recovered. 

\begin{figure}[htbp]
\centering
\includegraphics[width=.7\textwidth]{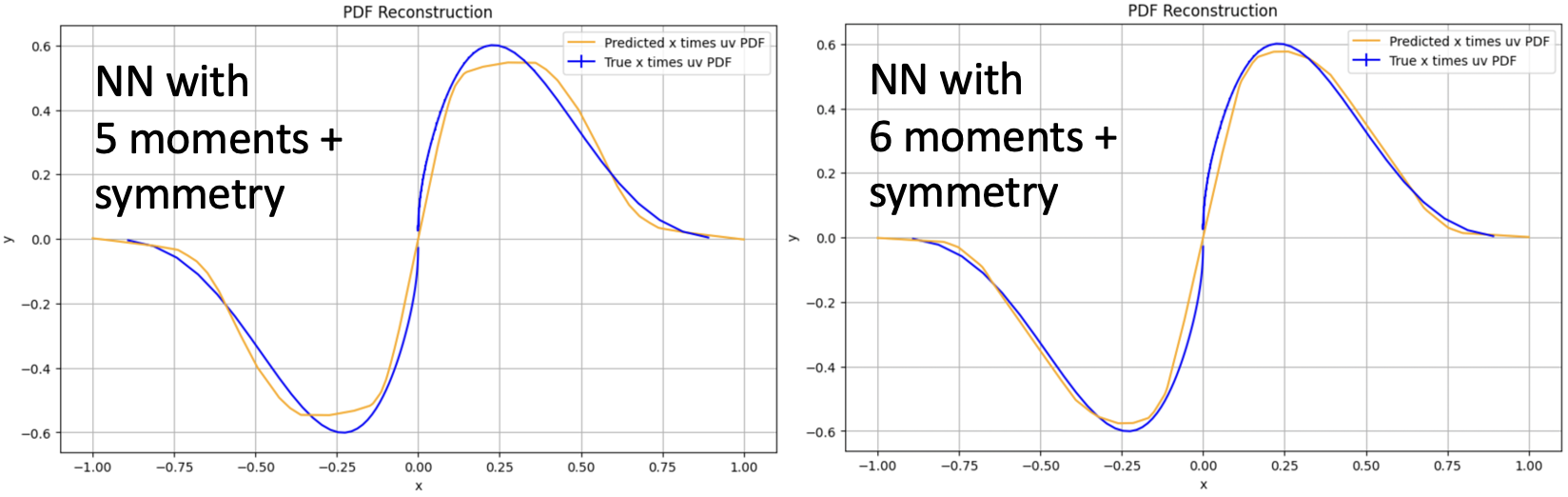}
\caption{PDF example of the application of the NNGPD framework. \label{fig:nn_first_ex}}
\end{figure}

\noindent We emphasize that the NNGPD approach within the framework of Figure \ref{fig:ml_framework}, and implementing the symmetries inherent in the description of GPDs, allow us to extract the $x$-dependence accurately with $n=5,6$ moments. In Figure \ref{fig:nn_first_ex} we show the working of the symmetries in the forward limit (PDF).






\section{Closure test on a GPD phenomenological model}

\noindent Before using the CFFs from experimental extractions and Mellin moments and GFFs from lattice calculations, a closure test needs to be performed on a realistic GPD model in order to benchmark the framework of Figure \ref{fig:ml_framework}. The parameterization used in this study, named UVA2, is from Ref.\cite{Panjsheeri:2025vpa}. UVA2 provides GPDs in the valence, sea, and gluon sectors that capture the behavior of lattice QCD moments, electromagnetic form factors, inclusive scattering data, and GPD perturbative QCD evolution equations. Additionally, UVA2 incorporates GPD symmetries in the DGLAP and ERBL regions and predicts CFFs in the kinematics of Jefferson Lab and the upcoming Electron-Ion Collider (EIC) at Brookhaven National Lab. 

\begin{figure}[htbp]
\centering
\includegraphics[width=1.0\textwidth]{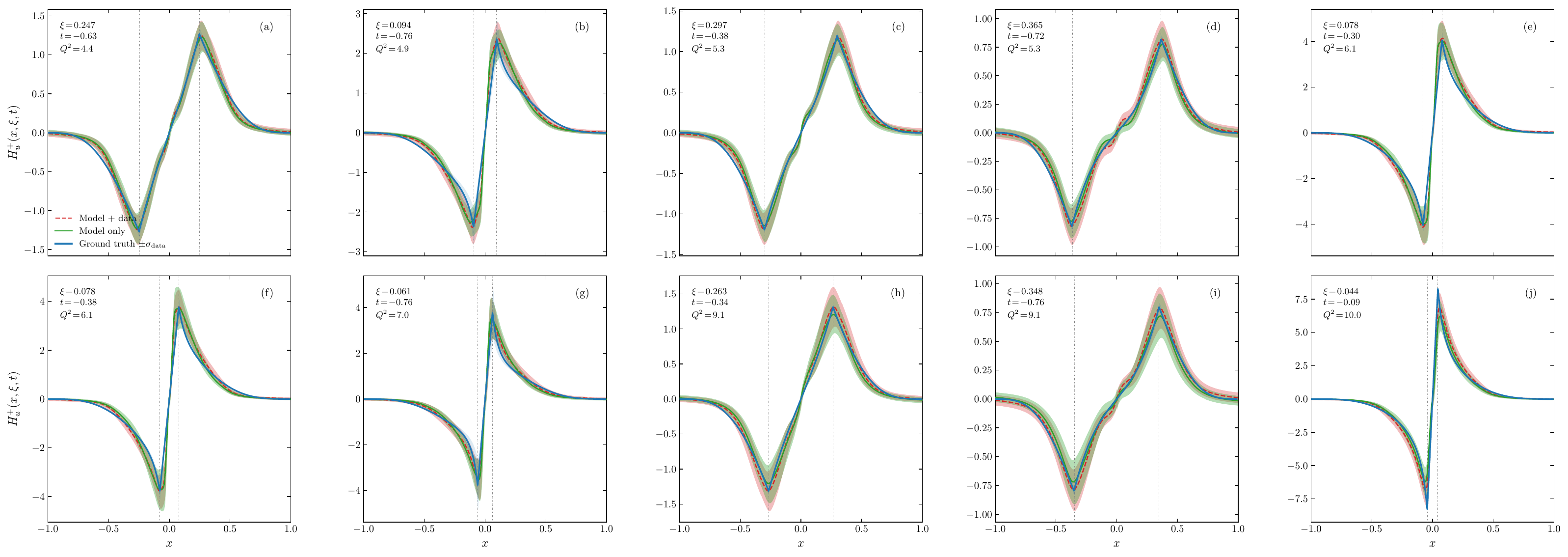}
\caption{(adapted from Ref.\cite{Xu:2026lko}) NNGPD  closure test for the $H^{+}_{u}$ GPD with a GPD parameterization, UVA2 \cite{Panjsheeri:2025vpa}. The uncertainties of UVA2 and NNGPD trained on UVA2 with and without the uncertainties of UVA2 are shown.
\label{fig:closure}
}
\end{figure}

\noindent With this parameterization in hand, a grid of CFFs and Mellin moments in a wide range of kinematics were produced that were then fed into the training of a neural network. Results for the antisymmetric $H_{u}^{+}$ GPD are shown in Figure \ref{fig:closure}. Results for the $H_{u}^{-}$, $H_{u}$ and $H_{\bar{u}}$ components are given in Ref.\cite{Xu:2026lko} using a BNN.

\section{Conclusions}

GPDs are complicated physical distributions that enter QCD in a manner which makes their extraction from experiment difficult, though not intractable. Machine learning techniques, as shown above within a closure test using the phenomenological UVA2 model, offer powerful methods for the extraction of GPDs from experimental data and ab-initio LQCD results coupled with theoretical constraints. Directly training the NNGPD deep learning framework with actual CFFs and lattice results remain as future work. Further enhancements of the NNGPD methodology will entail applying more interpretable tools, such as symbolic regression as in Ref. \cite{Dotson:2025omi}, to solving this inverse problem.



\acknowledgments

We acknowledge the many fruitful interactions with the ExclAIm collaboration. This work was supported by the U.S. Department of Energy under Grants DE-SC0024644 and DE-SC0016286. 





\bibliographystyle{unsrt}
\bibliography{biblio}





\end{document}